\documentclass[prl,twocolumn,showpacs,preprintnumbers,amsmath,amssymb]{revtex4}
\usepackage{colordvi}
\usepackage{amssymb}
\usepackage{graphics}
\usepackage{pstricks}
\usepackage{fancybox}
\usepackage{rotating}
\usepackage{epsfig}
\usepackage{subfigure}

\begin{document}
\preprint{CP3-09-04}

\title{$J/\psi$ production at HERA}

\author{P.~Artoisenet$^{a}$, J.~Campbell$^{b}$,  F.~Maltoni$^{a}$, F.~Tramontano$^{c}$}
\affiliation{
$^{a}$Center for Particle Physics and Phenomenology (CP3), Universit\'e catholique de Louvain, B-1348
Louvain-la-Neuve, Belgium\\
$^{b}$Department of Physics and Astronomy, University of Glasgow,
Glasgow G12 8QQ, United Kingdom \\
$^{c}$ Universit\`a di Napoli Federico II, Dipartimento di Scienze Fisiche, and INFN, Sezione di Napoli, I-80126 Napoli, Italy
}

%\date{}

\begin{abstract}

We consider the $J/\psi$ photo-production data collected at HERA in
the light of next-to-leading order predictions for the color-singlet
yield and polarization. We find that, while the shapes of inclusive distributions
in the transverse momentum and inelasticity are well reproduced,
the experimental rates are larger than those given by the
color-singlet contribution alone.  Furthermore, the  next-to-leading order
calculation predicts the  $J/\psi$'s to be mostly longitudinally polarized at
high transverse momentum in contrast with the trend of the preliminary
data from the ZEUS collaboration.

\end{abstract}


\maketitle

%\section{Introduction}

Non-relativistic QCD (NRQCD) provides a rigorous framework to
consistently separate high and low energy effects in quarkonium
production and decay~\cite{Bodwin:1994jh}. The factorization of the different energy scales
in the effective non-relativistic theory is achieved through an
expansion in $v$, the relative velocity of the heavy quarks in
the quarkonium state. If $v$ is a sufficiently small parameter, the
non-perturbative phenomena can be encoded into a few
process-independent long-distance matrix elements (LDMEs) which
can be extracted from experimental data.

Whether such an expansion is well-behaved for charmonium, and in particular
for the $J/\psi$,  is still subject to debate. So far no complete experimental evidence for the
universality of the LDMEs has been reached.
In particular, the role of the color-octet
transitions is not well assessed. Although these contributions seem to
be required to account for the observed $P_T$ spectrum in
hadro-production~\cite{Braaten:1994vv,Cacciari:1995yt}, they fail to
explain the polarization measurement at large transverse
momentum~\cite{Abulencia:2007us} and they do not seem to be needed to describe
fixed-target experiments~\cite{Maltoni:2006yp} and
photo-production~\cite{Adloff:2002ex,Chekanov:2002at} data.

Estimating the impact of the transitions at work for
$J/\psi$ production relies on the accuracy with which the corresponding
short-distance coefficients can be computed. Cross
sections at leading order in $\alpha_S$ are normally affected by
very large uncertainties and cannot give a reliable
estimate of the yield. In these cases, gathering information
on the underlying production mechanism(s) from data can be problematic.
The recent computation of next-to-leading order (NLO)  corrections in
$J/\psi$ electro-production~\cite{Zhang:2006ay,Zhang:2005cha,Gong:2009kp,Ma:2008gq} has
reduced the previous discrepancies between the NRQCD predictions and
the data. In hadro-production, higher-order $\alpha_S$ corrections to
the color-singlet production have been shown to significantly increase
the yield at large $P_T$~\cite{Campbell:2007ws,Gong:2008sn,Artoisenet:2008fc},
giving a better description of both the shape and the normalization of the $P_T$ spectrum.
Remarkably, the color-singlet
prediction for the polarization as a function of the $P_T$ also dramatically changes
once corrections in $\alpha_S$ are taken into account, leading to a better
agreement with the data.

In view of the recent theoretical progress related to $J/\psi$
hadro-production, it is worth reconsidering the photo-production data
obtained at HERA. Differential cross sections and polarization
observables for $J/\psi$ photo-production have been measured both by
the ZEUS and H1 collaborations.  Previous comparisons of these
measurements with the NRQCD predictions~\cite{Adloff:2002ex,Chekanov:2002at}
suggest that the  color-singlet yield alone reproduces the experimental
differential cross sections. However, theoretical uncertainties are
too large to draw any definite conclusion.

For the purpose of identifying the transitions that dominate
$J/\psi$ production, it is often useful to analyze the
polarization. Contrary to the differential cross sections,
one can expect the polarization predictions to be less
affected by the uncertainties associated with
the theoretical inputs. It has been argued~\cite{Beneke:1998re} that
the $J/\psi$ polarization could be a good discriminator between
color-singlet and color-octet transitions in $\gamma p$ collisions.
However, only leading order predictions have been considered so far
due to the lack of NLO calculations for the polarization for either
the singlet~\cite{Kramer:1995nb} or the octets~\cite{Maltoni:1997pt}.
Since QCD corrections can be very large in some regions of phase space,
especially for the color-singlet at large transverse
momentum, comparisons with leading order predictions are of  limited interest.

As new measurements of the $J/\psi$ polarization at HERA are still
being performed by the ZEUS and H1 collaborations, we limit ourselves
to a comparison with the preliminary data~\cite{Brugnera:2008DIS}.
For the color-singlet transition, polarization observables in photo-production
are given here at NLO accuracy in $\alpha_S$ for the first time.

The relevant kinematic variables for $J/\psi$ photoproduction are the
energy of the photon-proton system ($W$) and the fraction of the photon energy carried
by the $J/\psi$ in the proton rest frame ($z$),
\begin{equation}
W = \sqrt{(p_\gamma+p_p)^2}, \quad z=\frac{p_\psi.p_p}{p_\gamma.p_p} \;,
\end{equation}
as well as the transverse momentum of the $J/\psi$ ($P_T$).
Near the end-point region $z\simeq1$, $P_T\simeq0$ GeV,
$J/\psi$ production is enhanced by diffractive contributions
characterized by the exchange of colorless states. These contributions
cannot be correctly accounted for by the NRQCD factorization approach,
and hence they must be excluded from experimental measurements in order to make a
meaningful comparison. As diffractive production leads to a steeper slope for the $P_T$
spectrum of the $J/\psi$, the cut $P_T>1$ GeV effectively eliminates most of
the diffractive events. Whether this cut is enough to select a clean sample
of inelastic events is, however, still the subject of experimental investigation~\cite{Aktas:2003zi}. 
In the following, we adopt the cuts used by the ZEUS collaboration.

Inelastic contributions to $J/\psi$ production  include the
feed-down from excited charmonium states, which has not been taken into
account in this paper. Feed-down from $\psi(2S)$ is seen to lead to
a $15\%$ increase in the $J/\psi$ cross section~\cite{Chekanov:2002at},
whereas contributions from the decay of $\chi_c$ states are expected to be smaller ($\approx 1\%$).
The $J/\psi$ can also originate from a resolved photon. This is not a
technical problem since it requires a calculation
of hadro-production that is available~\cite{Campbell:2007ws}.
However, these contributions are only
significant in the low $z$ region, as we have checked by direct
computation. Since there is no data below $z =
0.3$, we consider only the direct contributions in this paper.

The ZEUS collaboration has published a measurement of the differential
cross sections for $J/\psi$ photo-production produced from electron-proton
collisions, with $E_p=820$ GeV and $E_e=27.5$ GeV~\cite{Chekanov:2002at}.
In their analysis, they consider the range $50 \textrm{ GeV }<W<180$
GeV, and convert electro-production
cross sections into photo-production cross sections by
dividing by the photon flux integrated in this range.  This is  estimated within
the  Weizs\"acker-Williams approximation: using $Q^{max}=1$ GeV, they
obtain  $F_{\textrm{photon}}=0.0987$ for the integrated photon flux.
For a consistent comparison with this measurement, our $\gamma p$
cross sections are convoluted with the photon flux using the
Weizs\"acker-Williams approximation, and then divided by the
integrated flux quoted above.

The cross section $\sigma(\gamma p \rightarrow J/\psi+X)$ in the
color-singlet model~\cite{Berger:1980ni} has 
been computed in Ref.~\cite{Kramer:1995nb} at NLO
accuracy in $\alpha_S$.  We reproduced this computation by using the
method and the results presented in Ref.~\cite{Campbell:2007ws} for the hadro-production
case, which also allows tracking of the $J/\psi$ polarization through
the angular correlations of the $J/\psi$ decay products, $J/\psi \to \ell^+ \ell^-$.
We have compared the results  for several inclusive observables, finding always
very good agreement.  As a further check and application of our results we also
calculated the NLO corrections to the decay width of a $^3S_1^{[1]}$
into $\gamma$ + hadrons. Our result is in excellent agreement with that of Ref.~\cite{Kramer:1999bf}.

\begin{table}[t]
$$
\begin{array}{c|ccc}
\hline\hline
\mu_r \backslash  \mu_f & 0.5 \, \mu_0 &  \mu_0  & 2 \,  \mu_0  \\
\hline
2 \, \mu_0 & 8.07  & 9.67  & 10.6  \\
 \mu_0 & 9.45  & 10.2  & 10.3  \\
0.5 \, \mu_0 & 10.6  & 8.28 & 5.96   \\
\hline\hline
\end{array}
$$
\caption{\label{tab1}
Scale dependence of the total cross section $\sigma(P_T>1 \textrm{ GeV})$  (expressed in nb).
$W$ is set to $100$ GeV. Other input parameters are explained in the text.}
\end{table}

\begin{figure}[h]
\centering
\includegraphics[scale=.65]{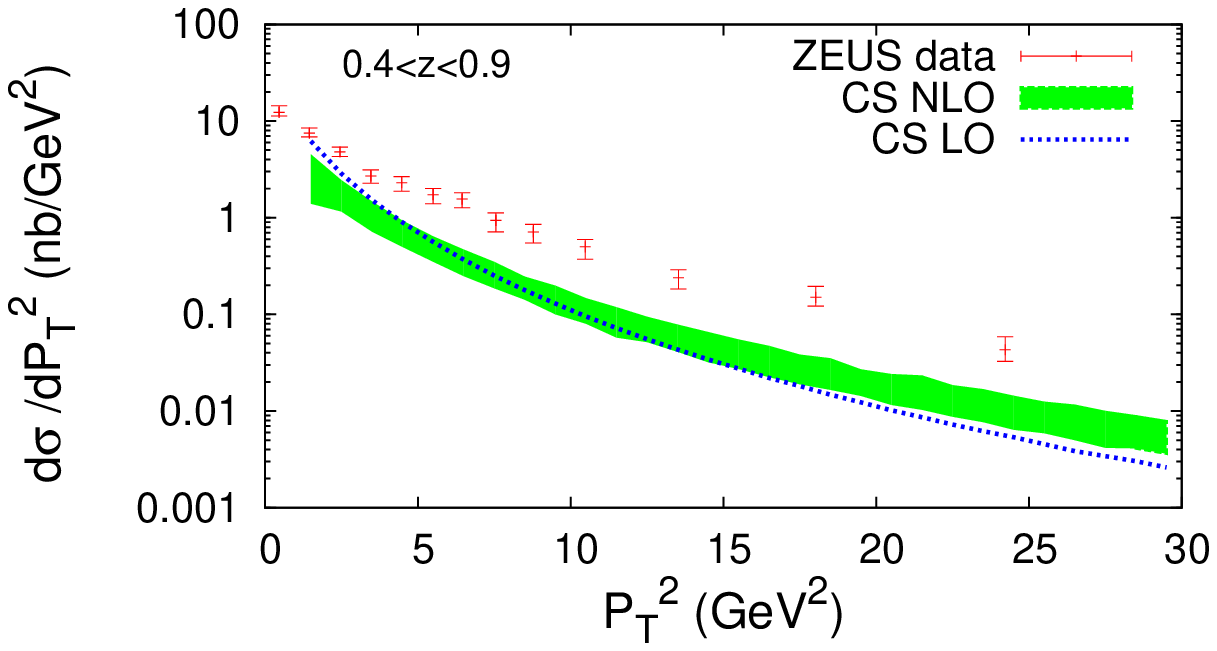}
\includegraphics[scale=.65]{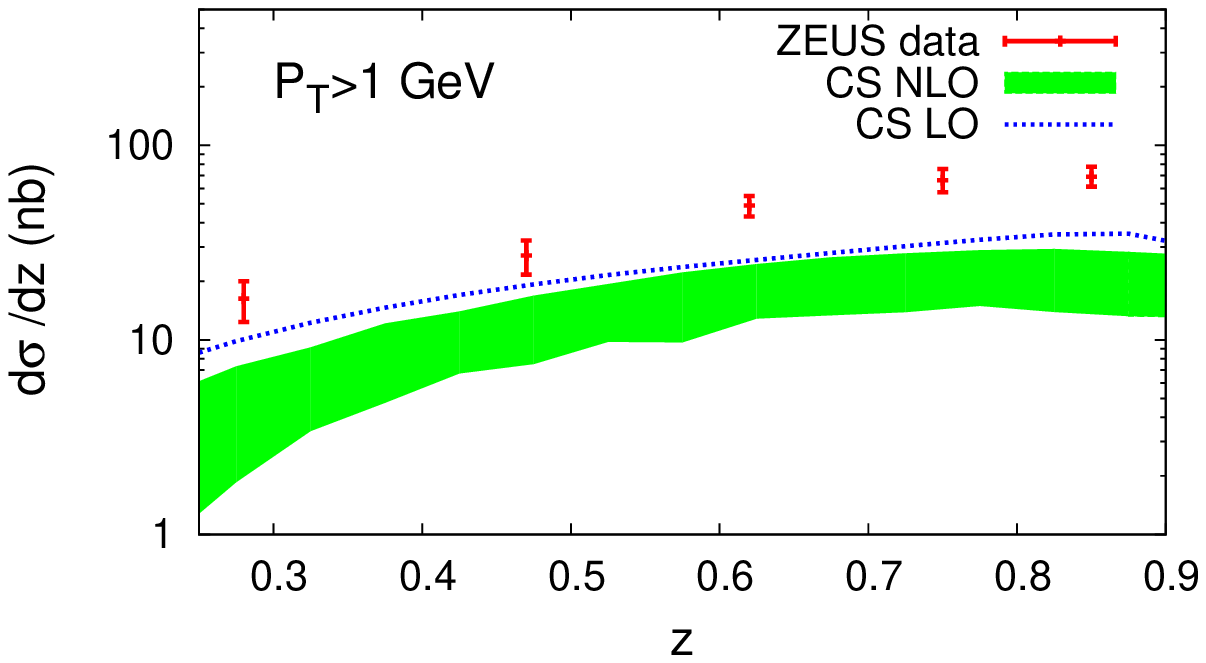}
\caption{Differential cross sections in $P_T^2$
and  $z$, compared with the ZEUS data~\cite{Chekanov:2002at}.}
\label{diff_crossX}
\end{figure}

In our photo-production analysis,  we fix $\langle
\mathcal{O}_{J/\psi}(^3S_1[1]) \rangle=1.16$ GeV$^3$ and we use the
CTEQ6M pdf set~\cite{Pumplin:2002vw}.
Given the small range in $p_T$ available and for the sake of
simplicity we prefer to use fixed scales.
As the central value we choose $\mu_0=4m_c$, i.e.,
where the sensitivity of the cross section 
$\sigma(P_T>1 \textrm{ GeV})$ to scale variations
is minimal (see Table \ref{tab1}).
Since a large variation is observed
when we vary the scales in opposite directions, we impose
the additional condition $0.5<\frac{\mu_r}{\mu_f}<2$
in our prediction for the differential cross sections.
We also vary the charm quark mass in the range $1.4-1.6$ GeV.
Scale and mass uncertainties are combined in quadrature.
The resulting distributions are displayed in Fig. \ref{diff_crossX}.

For comparison, we also plot the color-singlet prediction at leading
order in $\alpha_S$, for which we use the CTEQ6L1 pdf set.
As can be seen from Fig.~\ref{diff_crossX}, the $\alpha_S$
corrections increase the differential cross section
in the high $P_T$ region, where the yield is dominated by the new
channels that open up at order $\alpha_S^3$. Nevertheless, the
color-singlet yield at NLO clearly undershoots the ZEUS data.  The plots
in Fig.~\ref{diff_crossX} differ from the comparison presented in
Ref.~\cite{Chekanov:2002at}, where rather extreme values for the
renormalization scale were used that have the effect of artificially
increasing the normalization. We do not display here the comparison
with the $P_T^2$ and $z$ distributions measured by the H1
collaboration~\cite{Adloff:2002ex} since it shares the same features.

We now turn to the polarization. Experimentally, the polarization of
the $J/\psi$'s can be traced back by analyzing the angular distribution
of the leptons originating from the decay of the $J/\psi$. It is convenient to
decompose this angular distribution in terms of the polar and
azimuthal angles $\theta$ and $\phi$ in the $J/\psi$ rest frame:
\begin{eqnarray}
\frac{d \sigma}{d\Omega dy}  & \propto &  1+\lambda(y) \cos^2  \nonumber  \theta +\mu(y)\sin 2\theta \cos\phi \\ & & + \frac{\nu(y)}{2} \sin^2\theta \cos 2\phi
\end{eqnarray}
where $y$ stands for a certain (set of) variable(s) (either $P_T$ or $z$ in the following).
If the polar axis coincides with the spin quantization axis,
the parameters $\lambda$, $\mu$, $\nu$ can be related to the spin density matrix
elements:
\begin{equation}
\lambda=\frac{\rho_{1,1}-\rho_{0,0}}{\rho_{1,1}+\rho_{0,0}}, \quad \mu=\frac{\sqrt 2 Re \rho_{1,0}}{\rho_{1,1}+\rho_{0,0}}, \quad \nu=\frac{2 \rho_{1,-1}}{\rho_{1,1}+\rho_{0,0}}.
\label{pol_params}
\end{equation}

The spin information that we extract in this way depends
on the choice of the quantization axis. Here, we  decide
to work in the target frame ($\hat {\boldsymbol z}=-\frac{\boldsymbol{p}_{\textrm{p}}}{|\boldsymbol p_{\textrm{p}}|}$) as in the recent analysis performed by the ZEUS collaboration.

As mentioned above, in our NLO computation the $J/\psi$
decays into leptons.  In order to obtain $\lambda$ (resp. $\nu$) we
have integrated the cross sections over $\phi$ (resp. $\theta$)
and extracted the polarization parameters by fitting the resulting distributions.
In so doing, we have used the same kinematic conditions as those considered by the ZEUS collaboration:
$E_p=920$ GeV, $E_e=27.5$ GeV, $P_T>1$ GeV,  $z>0.4$ and $\quad 50 \textrm{ GeV}<W<
180$ GeV.
Notice that the cut $z<0.9$ previously considered for the differential
cross sections has been relaxed here.  The measurement of the
polarization versus $P_T$ is therefore subject to a larger
contamination from diffractive contributions.

\begin{figure}[h!]
\centering
\includegraphics[scale=.65]{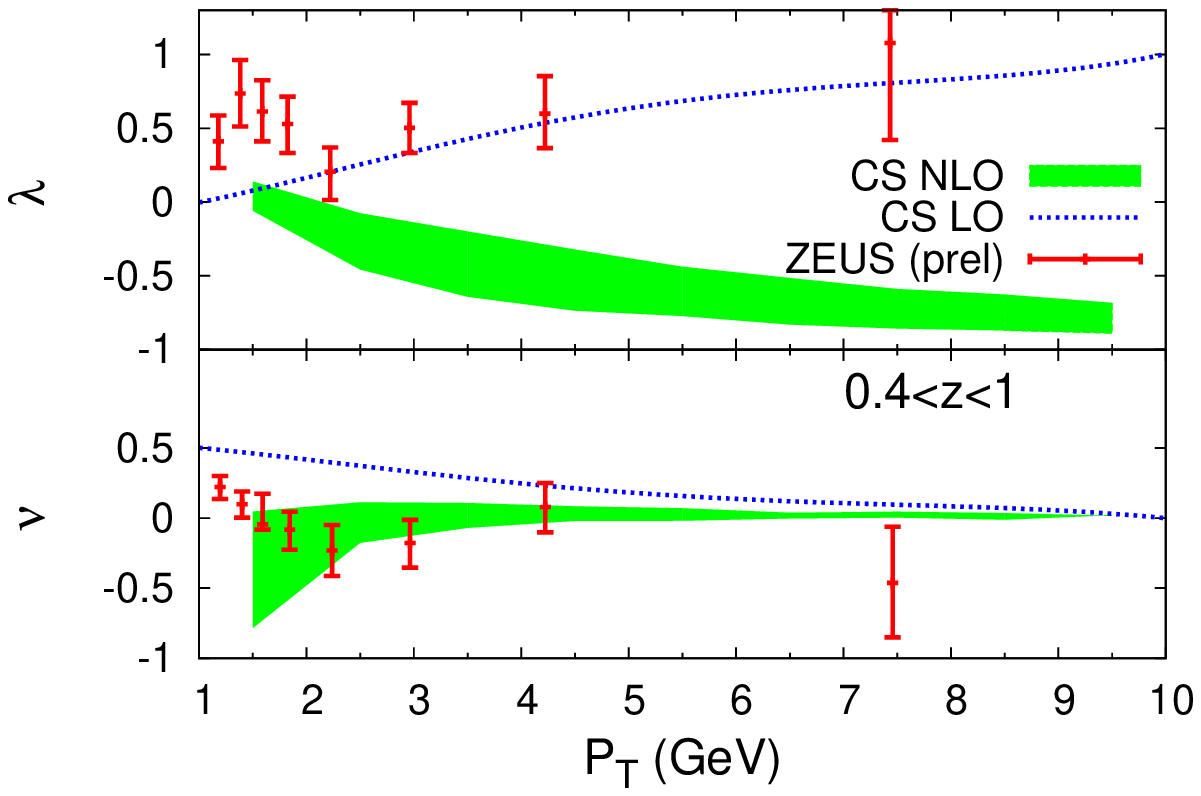}
\vspace{-0.5cm}
\includegraphics[scale=.65]{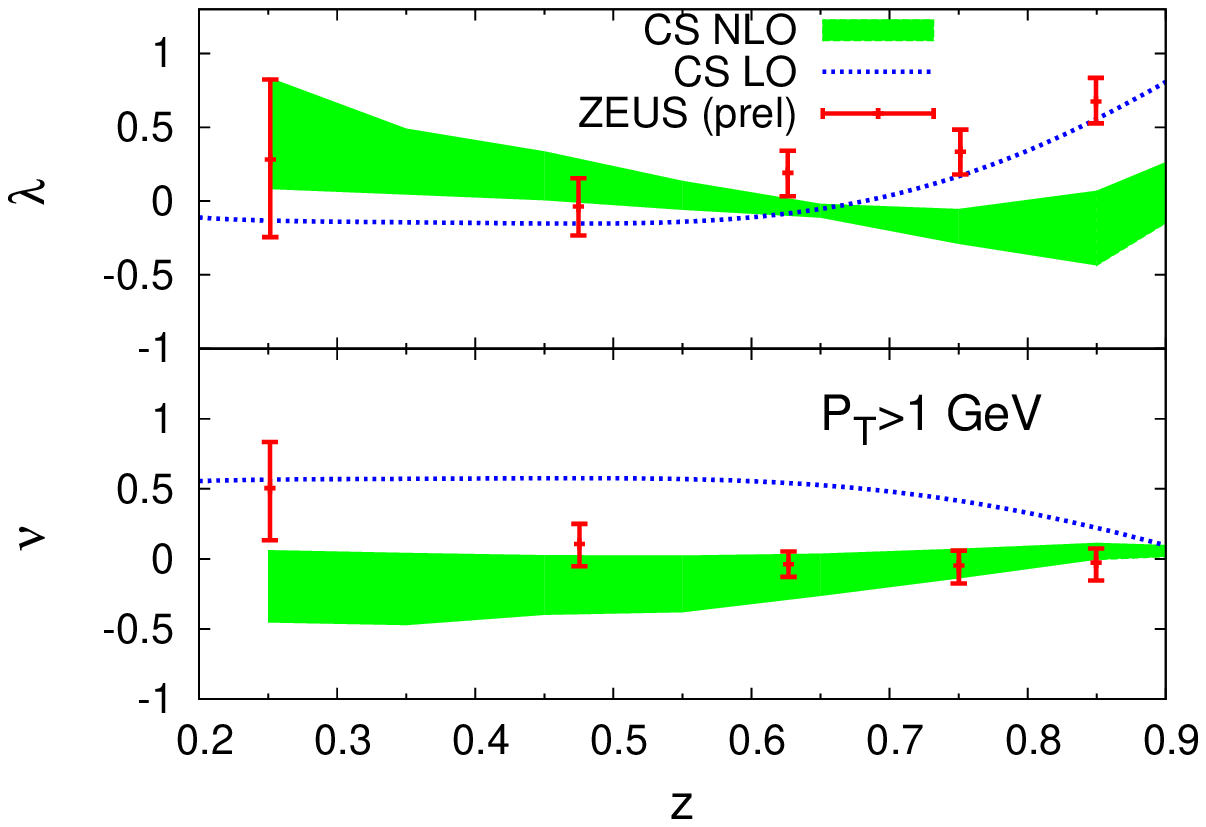}
\caption{Polarization parameters for color-singlet production
at leading order and next-to-leading order in $\alpha_S$,
compared with the ZEUS preliminary measurement.}
\label{polarization}
\end{figure}

The NLO predictions of the polarization parameters associated with
color-singlet production are displayed in Fig.~\ref{polarization}, together with
the LO predictions and the preliminary ZEUS measurements.
The band comes from the uncertainties associated
with the choice of the scales -- varied in the range defined by
$0.5 \mu_0<\mu_f, \mu_r<2 \mu_0$ and
$0.5<\frac{\mu_r}{\mu_f}<2$ -- which is much larger
than the mass uncertainty. For some specific values of the scales
(namely $\mu_r=0.5$), the $\lambda$ and $\nu$ parameters appear
to be unphysical in some bins. This is due to the fact
that in these cases our calculation leads to a negative value for the
diagonal components of the spin density matrix
at $P_T \lesssim 1$ GeV, and hence cannot be trusted in this region.
Therefore, in the error bands in Fig.~\ref{polarization},
we have disregarded scale choices leading to unphysical predictions.

QCD corrections to the color-singlet production have a strong impact on the
polarization prediction.  The most spectacular effect comes from the
behavior of the $\lambda$ parameter at large transverse momentum, for
which the prediction is rather stable under variation of the scales.
At leading order in $\alpha_S$, the color-singlet transition gives a
transverse $J/\psi$ at large $P_T$.  Once  QCD corrections are included,
the $\lambda$ parameter decreases rapidly and has a large
negative value above $P_T=4-5$ GeV.  This situation is similar to that
in hadro-production where the $J/\psi$ produced via a color-singlet transition
is longitudinal at large transverse momentum~\cite{Gong:2008sn,Artoisenet:2008fc}.  Such a correction
for the $\lambda$ parameter at moderate and high $P_T$, as well as
the decrease at $z\approx 0.8$, is not supported by the preliminary data
from the ZEUS collaboration and therefore suggests the presence of other mechanisms for $J/\psi$ production.
In the low $z$ region, the  scale uncertainty is too large to draw any
conclusion. QCD corrections to color-singlet production also affect the value of the
$\nu$ parameter, which is closer to the experimental data in comparison with the prediction at leading order.

In this letter, we have studied $J/\psi$ photo-production via a
color-singlet transition at HERA and presented for the first time
a comparison with the predicted polarization at NLO accuracy.
Taking into account the $\alpha_S$ corrections, color-singlet
production alone does not describe all features of the data collected at HERA. With a
natural choice for the renormalization scale, the predicted rate
is smaller than data, even though the shapes of the differential distributions
are well described. Moreover, the preliminary measurement of the $J/\psi$
polarization by the ZEUS collaboration as a function of the $P_T$ shows
a very different trend with respect to the theoretical predictions.

New channels appearing at NNLO in $\alpha_S$ (including fragmentation processes)
may increase the differential cross section, as has been pointed
out in the hadroproduction case~\cite{Artoisenet:2008fc}.
Although the kinematics differ in photoproduction,
such contributions could give an enhancement
at large transverse momentum ($P_T>5$ GeV).  However, no
reliable  estimate of NNLO  contributions in the $P_T$ region
accessible  at HERA is presently available.

The current discrepancies could possibly be solved by invoking
color-octet transitions, i.e. contributions from the intermediate
states $^1S_0^{[8]}$ and $^3P_J^{[8]}$. Unfortunately, any
phenomenological analysis of the impact of these contributions on
differential cross sections and polarization observables is limited by
the omission of higher-order corrections that are currently unknown. A
complete $\alpha_S^3$ computation, particularly for the prediction of
the polarization of the $J/\psi$ produced via a $P-$wave color-octet
state, would be welcome in order to shed further light on the mechanisms at
work in photo-production.

We are thankful to Michael Kr\"amer for the accurate comparison
with the calculation of Ref.~\cite{Kramer:1995nb} and for numerous discussions.
We also thank Alessandro Bertolin and Riccardo Brugnera for their help
with the ZEUS data. This work was supported by the Fonds National de la Recherche
Scientiﬁque and by the Belgian Federal Office for Scientific, Technical and Cultural
Affairs through the Interuniversity Attraction Pole No. P6/11.

\bibliography{database}

\begin{thebibliography}{22}
\expandafter\ifx\csname natexlab\endcsname\relax\def\natexlab#1{#1}\fi
\expandafter\ifx\csname bibnamefont\endcsname\relax
  \def\bibnamefont#1{#1}\fi
\expandafter\ifx\csname bibfnamefont\endcsname\relax
  \def\bibfnamefont#1{#1}\fi
\expandafter\ifx\csname citenamefont\endcsname\relax
  \def\citenamefont#1{#1}\fi
\expandafter\ifx\csname url\endcsname\relax
  \def\url#1{\texttt{#1}}\fi
\expandafter\ifx\csname urlprefix\endcsname\relax\def\urlprefix{URL }\fi
\providecommand{\bibinfo}[2]{#2}
\providecommand{\eprint}[2][]{\url{#2}}

\bibitem[{\citenamefont{Bodwin et~al.}(1995)\citenamefont{Bodwin, Braaten, and
  Lepage}}]{Bodwin:1994jh}
\bibinfo{author}{\bibfnamefont{G.~T.} \bibnamefont{Bodwin}},
  \bibinfo{author}{\bibfnamefont{E.}~\bibnamefont{Braaten}}, \bibnamefont{and}
  \bibinfo{author}{\bibfnamefont{G.~P.} \bibnamefont{Lepage}},
  \bibinfo{journal}{Phys. Rev.} \textbf{\bibinfo{volume}{D51}},
  \bibinfo{pages}{1125} (\bibinfo{year}{1995}), \eprint{hep-ph/9407339}.

\bibitem[{\citenamefont{Braaten and Fleming}(1995)}]{Braaten:1994vv}
\bibinfo{author}{\bibfnamefont{E.}~\bibnamefont{Braaten}} \bibnamefont{and}
  \bibinfo{author}{\bibfnamefont{S.}~\bibnamefont{Fleming}},
  \bibinfo{journal}{Phys. Rev. Lett.} \textbf{\bibinfo{volume}{74}},
  \bibinfo{pages}{3327} (\bibinfo{year}{1995}), \eprint{hep-ph/9411365}.

\bibitem[{\citenamefont{Cacciari et~al.}(1995)\citenamefont{Cacciari, Greco,
  Mangano, and Petrelli}}]{Cacciari:1995yt}
\bibinfo{author}{\bibfnamefont{M.}~\bibnamefont{Cacciari}},
  \bibinfo{author}{\bibfnamefont{M.}~\bibnamefont{Greco}},
  \bibinfo{author}{\bibfnamefont{M.~L.} \bibnamefont{Mangano}},
  \bibnamefont{and} \bibinfo{author}{\bibfnamefont{A.}~\bibnamefont{Petrelli}},
  \bibinfo{journal}{Phys. Lett.} \textbf{\bibinfo{volume}{B356}},
  \bibinfo{pages}{553} (\bibinfo{year}{1995}), \eprint{hep-ph/9505379}.

\bibitem[{\citenamefont{Abulencia et~al.}(2007)}]{Abulencia:2007us}
\bibinfo{author}{\bibfnamefont{A.}~\bibnamefont{Abulencia}}
  \bibnamefont{et~al.} (\bibinfo{collaboration}{CDF}), \bibinfo{journal}{Phys.
  Rev. Lett.} \textbf{\bibinfo{volume}{99}}, \bibinfo{pages}{132001}
  (\bibinfo{year}{2007}), \eprint{0704.0638}.

\bibitem[{\citenamefont{Maltoni et~al.}(2006)}]{Maltoni:2006yp}
\bibinfo{author}{\bibfnamefont{F.}~\bibnamefont{Maltoni}} \bibnamefont{et~al.},
  \bibinfo{journal}{Phys. Lett.} \textbf{\bibinfo{volume}{B638}},
  \bibinfo{pages}{202} (\bibinfo{year}{2006}), \eprint{hep-ph/0601203}.

\bibitem[{\citenamefont{Adloff et~al.}(2002)}]{Adloff:2002ex}
\bibinfo{author}{\bibfnamefont{C.}~\bibnamefont{Adloff}} \bibnamefont{et~al.}
  (\bibinfo{collaboration}{H1}), \bibinfo{journal}{Eur. Phys. J.}
  \textbf{\bibinfo{volume}{C25}}, \bibinfo{pages}{25} (\bibinfo{year}{2002}),
  \eprint{hep-ex/0205064}.

\bibitem[{\citenamefont{Chekanov et~al.}(2003)}]{Chekanov:2002at}
\bibinfo{author}{\bibfnamefont{S.}~\bibnamefont{Chekanov}} \bibnamefont{et~al.}
  (\bibinfo{collaboration}{ZEUS}), \bibinfo{journal}{Eur. Phys. J.}
  \textbf{\bibinfo{volume}{C27}}, \bibinfo{pages}{173} (\bibinfo{year}{2003}),
  \eprint{hep-ex/0211011}.

\bibitem[{\citenamefont{Zhang and Chao}(2007)}]{Zhang:2006ay}
\bibinfo{author}{\bibfnamefont{Y.-J.} \bibnamefont{Zhang}} \bibnamefont{and}
  \bibinfo{author}{\bibfnamefont{K.-T.} \bibnamefont{Chao}},
  \bibinfo{journal}{Phys. Rev. Lett.} \textbf{\bibinfo{volume}{98}},
  \bibinfo{pages}{092003} (\bibinfo{year}{2007}), \eprint{hep-ph/0611086}.

\bibitem[{\citenamefont{Zhang et~al.}(2006)\citenamefont{Zhang, Gao, and
  Chao}}]{Zhang:2005cha}
\bibinfo{author}{\bibfnamefont{Y.-J.} \bibnamefont{Zhang}},
  \bibinfo{author}{\bibfnamefont{Y.-J.} \bibnamefont{Gao}}, \bibnamefont{and}
  \bibinfo{author}{\bibfnamefont{K.-T.} \bibnamefont{Chao}},
  \bibinfo{journal}{Phys. Rev. Lett.} \textbf{\bibinfo{volume}{96}},
  \bibinfo{pages}{092001} (\bibinfo{year}{2006}), \eprint{hep-ph/0506076}.

\bibitem[{\citenamefont{Gong and Wang}(2009)}]{Gong:2009kp}
\bibinfo{author}{\bibfnamefont{B.}~\bibnamefont{Gong}} \bibnamefont{and}
  \bibinfo{author}{\bibfnamefont{J.-X.} \bibnamefont{Wang}}
  (\bibinfo{year}{2009}), \eprint{0901.0117}.

\bibitem[{\citenamefont{Ma et~al.}(2008)\citenamefont{Ma, Zhang, and
  Chao}}]{Ma:2008gq}
\bibinfo{author}{\bibfnamefont{Y.-Q.} \bibnamefont{Ma}},
  \bibinfo{author}{\bibfnamefont{Y.-J.} \bibnamefont{Zhang}}, \bibnamefont{and}
  \bibinfo{author}{\bibfnamefont{K.-T.} \bibnamefont{Chao}}
  (\bibinfo{year}{2008}), \eprint{0812.5106}.

\bibitem[{\citenamefont{Campbell et~al.}(2007)\citenamefont{Campbell, Maltoni,
  and Tramontano}}]{Campbell:2007ws}
\bibinfo{author}{\bibfnamefont{J.}~\bibnamefont{Campbell}},
  \bibinfo{author}{\bibfnamefont{F.}~\bibnamefont{Maltoni}}, \bibnamefont{and}
  \bibinfo{author}{\bibfnamefont{F.}~\bibnamefont{Tramontano}},
  \bibinfo{journal}{Phys. Rev. Lett.} \textbf{\bibinfo{volume}{98}},
  \bibinfo{pages}{252002} (\bibinfo{year}{2007}), \eprint{hep-ph/0703113}.

\bibitem[{\citenamefont{Gong and Wang}(2008)}]{Gong:2008sn}
\bibinfo{author}{\bibfnamefont{B.}~\bibnamefont{Gong}} \bibnamefont{and}
  \bibinfo{author}{\bibfnamefont{J.-X.} \bibnamefont{Wang}}
  (\bibinfo{year}{2008}), \eprint{0802.3727}.

\bibitem[{\citenamefont{Artoisenet et~al.}(2008)\citenamefont{Artoisenet,
  Campbell, Lansberg, Maltoni, and Tramontano}}]{Artoisenet:2008fc}
\bibinfo{author}{\bibfnamefont{P.}~\bibnamefont{Artoisenet}},
  \bibinfo{author}{\bibfnamefont{J.}~\bibnamefont{Campbell}},
  \bibinfo{author}{\bibfnamefont{J.~P.} \bibnamefont{Lansberg}},
  \bibinfo{author}{\bibfnamefont{F.}~\bibnamefont{Maltoni}}, \bibnamefont{and}
  \bibinfo{author}{\bibfnamefont{F.}~\bibnamefont{Tramontano}},
  \bibinfo{journal}{Phys. Rev. Lett.} \textbf{\bibinfo{volume}{101}},
  \bibinfo{pages}{152001} (\bibinfo{year}{2008}), \eprint{0806.3282}.

\bibitem[{\citenamefont{Beneke et~al.}(1998)\citenamefont{Beneke, Kramer, and
  Vanttinen}}]{Beneke:1998re}
\bibinfo{author}{\bibfnamefont{M.}~\bibnamefont{Beneke}},
  \bibinfo{author}{\bibfnamefont{M.}~\bibnamefont{Kramer}}, \bibnamefont{and}
  \bibinfo{author}{\bibfnamefont{M.}~\bibnamefont{Vanttinen}},
  \bibinfo{journal}{Phys. Rev.} \textbf{\bibinfo{volume}{D57}},
  \bibinfo{pages}{4258} (\bibinfo{year}{1998}), \eprint{hep-ph/9709376}.

\bibitem[{\citenamefont{Kramer}(1996)}]{Kramer:1995nb}
\bibinfo{author}{\bibfnamefont{M.}~\bibnamefont{Kramer}},
  \bibinfo{journal}{Nucl. Phys.} \textbf{\bibinfo{volume}{B459}},
  \bibinfo{pages}{3} (\bibinfo{year}{1996}), \eprint{hep-ph/9508409}.

\bibitem[{\citenamefont{Maltoni et~al.}(1998)\citenamefont{Maltoni, Mangano,
  and Petrelli}}]{Maltoni:1997pt}
\bibinfo{author}{\bibfnamefont{F.}~\bibnamefont{Maltoni}},
  \bibinfo{author}{\bibfnamefont{M.~L.} \bibnamefont{Mangano}},
  \bibnamefont{and} \bibinfo{author}{\bibfnamefont{A.}~\bibnamefont{Petrelli}},
  \bibinfo{journal}{Nucl. Phys.} \textbf{\bibinfo{volume}{B519}},
  \bibinfo{pages}{361} (\bibinfo{year}{1998}), \eprint{hep-ph/9708349}.

\bibitem[{\citenamefont{Brugnera}(2008)}]{Brugnera:2008DIS}
\bibinfo{author}{\bibfnamefont{R.}~\bibnamefont{Brugnera}}
  (\bibinfo{year}{2008}), \bibinfo{note}{proc.~of XVI Int.~Workshop on
  Deep-Inelastic Scattering and Related Topics, London, England, April 2008}.

\bibitem[{\citenamefont{Aktas et~al.}(2003)}]{Aktas:2003zi}
\bibinfo{author}{\bibfnamefont{A.}~\bibnamefont{Aktas}} \bibnamefont{et~al.}
  (\bibinfo{collaboration}{H1}), \bibinfo{journal}{Phys. Lett.}
  \textbf{\bibinfo{volume}{B568}}, \bibinfo{pages}{205} (\bibinfo{year}{2003}),
  \eprint{hep-ex/0306013}.

\bibitem[{\citenamefont{Berger and Jones}(1981)}]{Berger:1980ni}
\bibinfo{author}{\bibfnamefont{E.~L.} \bibnamefont{Berger}} \bibnamefont{and}
  \bibinfo{author}{\bibfnamefont{D.~L.} \bibnamefont{Jones}},
  \bibinfo{journal}{Phys. Rev.} \textbf{\bibinfo{volume}{D23}},
  \bibinfo{pages}{1521} (\bibinfo{year}{1981}).

\bibitem[{\citenamefont{Kramer}(1999)}]{Kramer:1999bf}
\bibinfo{author}{\bibfnamefont{M.}~\bibnamefont{Kramer}},
  \bibinfo{journal}{Phys. Rev.} \textbf{\bibinfo{volume}{D60}},
  \bibinfo{pages}{111503(R)} (\bibinfo{year}{1999}), \eprint{hep-ph/9904416}.

\bibitem[{\citenamefont{Pumplin et~al.}(2002)}]{Pumplin:2002vw}
\bibinfo{author}{\bibfnamefont{J.}~\bibnamefont{Pumplin}} \bibnamefont{et~al.},
  \bibinfo{journal}{JHEP} \textbf{\bibinfo{volume}{07}}, \bibinfo{pages}{012}
  (\bibinfo{year}{2002}), \eprint{hep-ph/0201195}.

\end{thebibliography}

\end{document}